\documentstyle[aps,twocolumn,epsfig]{revtex}

\newcommand{\beq}{\begin{equation}}
\newcommand{\eeq}{\end{equation}}
\newcommand{\bea}{\begin{eqnarray}}
\newcommand{\eea}{\end{eqnarray}}
\newcommand{\ba}{\begin{array}}
\newcommand{\ea}{\end{array}}
\newcommand{\bc}{\begin{center}}
\newcommand{\ec}{\end{center}}

\newcommand{\ibid}{{\it ibid. }}
\newcommand{\bml}{\begin{mathletters}}
\newcommand{\eml}{\end{mathletters}}
\newcommand{\commentout}[1]{{}}

\newcommand{\lanl}{LANL e-print}

\begin{document}
\wideabs{
\title{Landau-Zener Problem for Trilinear Hamiltonians}
\author{Artur Ishkhanyan}
\address{Engineering Center of Armenian National
Academy of Sciences,\\
Ashtarak-2, 378410 Armenia}
\author{Matt Mackie}
\address{Helsinki Institute of Physics, University of
Helsinki, PL 64,
\\
FIN-00014 Helsingin yliopisto, Finland}
\author{Andrew Carmichael, Phillip L. Gould, and Juha
Javanainen}
\address{Department of Physics, University of
Connecticut, Storrs,\\
Connecticut 06269-3046}
\date{\today }
\maketitle

\begin{abstract}
We consider a nonlinear version of the Landau-Zener problem, focusing on
photoassociation of a Bose-Einstein  condensate as
a specific example. Contrary to the exponential rate dependence obtained
for the linear problem, a series expansion technique indicates that, when
the resonance is crossed slowly, the probability for failure of
adiabaticity is directly proportional to the rate at which the resonance
is crossed.
\end{abstract}
}

When external conditions force the diabatic energies
of two interacting
quantum states cross, under conditions of complete
adiabaticity the
system
makes a transition from one bare energy eigenstate to
the other.
However,
``complete adiabaticity'' applies when the levels
cross ``infinitely
slowly''. The realistic case when only a finite time
is spent near
resonance makes the Landau-Zener (LZ)
problem~\cite{LZSTUFF1,LZSTUFF2},
one of the handful of key paradigms in quantum
mechanics. On the other
hand, the Gross-Pitaevskii equation and its variants,
nonlinear
versions
of standard quantum mechanics, have proven highly
successful in the
theoretical description of alkali-vapor Bose-Einstein
condensate~\cite{DAL99}. Nonlinear quantum mechanics
gives up
the superposition principle and hence the linear
vector space
structure.
How much of our intuitive understanding of quantum
mechanics survives
becomes an issue.

The nonlinear version of the LZ problem that is the
focus of this Letter is
something we have come across in our studies of
photoassociation~\cite{JAV99,KOS00} of a Bose-Einstein
condensate
(BEC). However, the same question will come up in an
attempt to control the
scattering length~\cite{ROB01} of an atomic BEC by
means of a Feshbach
resonance~\cite{TIM99}, in second-harmonic generation
in nonlinear
optics~\cite{WAL72}, and generally in field theories
where the nonlinearity is of
the
form $\psi ^{\dagger }\phi \phi $. For concreteness we
couch our
discussion in terms of photoassociation, and thus
consider processes in which
two atoms and a laser photon join to make a molecule.
It is known from
a
second-quantized numerical solution that by sweeping
the frequency of
the
photoassociating laser, in the adiabatic limit it is
possible to
convert
an atomic BEC entirely into a molecular BEC~\cite
{JAV99}. The question is, what are the nonlinear consequences if the
frequency is not
swept ``infinitely slowly''? While we
actually know of no publication addressing this
question~\cite{MIE00,YUR01}, we have
found that the same applies also to the semiclassical
version of the problem,
when the atomic and molecular condensates are
described by a nonlinear
two-state system. 

The answer to the above question is outlined as follows. First
the
essentials of the linear LZ problem are reviewed,
which serves both as an
introduction to the formalism and a foundation for our
series expansion. Turning to
the nonlinear LZ model, we use photoassociation as an
example to review the
dressed state picture and illustrate the physics of
adiabatic following. In particular,
a picture in terms of crossing of stationary states
still applies and predicts
near-adiabaticity, but the details are qualitatively
different from the case of linear
quantum mechanics. The first-order Heisenberg
equations of
motion for the semi-classical atom-molecule amplitudes
are then converted into a
third-order differential equation for the molecular
probability, whereby a
perturbative expansion allows for what amounts to a
Green function solution. All
told, we find that, while in the linear case the
probability for no transition is an
exponential function of the speed  at which the
resonance is crossed, at slow sweep
rates of the laser frequency the probability that an
atom does not make a transition
to  the molecular condensate is directly proportional
to the rate. Before closing, we estimate explicit numbers for two
systems, $^{23}$Na and $^{87}$Rb, that could serve to test our prediction.

The linear LZ problem is specified by the equations
\begin{equation}
i\dot{\alpha}=\Omega \beta ,\qquad i\dot{\beta}=\Omega
\alpha +\delta \beta
\,.  \label{LINLZ}
\end{equation}
Here $\alpha $ and $\beta $ are the probability
amplitudes
for the two bare or diabatic states, $\Omega $ is the
(real) coupling
between them, and the detuning $\delta =\xi t$ stands
for the time
dependent frequency difference between the two states.
For any fixed
$\delta $ the system~(\ref{LINLZ}) has two
nondegenerate
dressed~\cite{CTTEXT} or adiabatic states
$[\bar{\alpha},
\bar{\beta}]$ with the property that the time
dependent solution is of
the
form $[\alpha (t),\beta (t)]=e^{-i\epsilon
t}[\bar{\alpha},\bar{\beta}_0]$
for some quasienergy $\hbar \epsilon $. In particular,
one of these
states
is such that $[\bar{\alpha},\bar{\beta}]\rightarrow
[1,0]$ when $\delta
\rightarrow -\infty $, and evolves continuously to
$[0,1]$ as $\delta
\rightarrow \infty $.

Suppose that the system with the time dependent
detuning $\delta =\xi
t$
starts out in the state with $\alpha =1$ when
$t\rightarrow -\infty $,
then
it also start out in this dressed state. As the
detuning is swept
``infinitely slowly'' to $+\infty $, by the adiabatic
theorem the
system
emerges at time $t=\infty $ in the same dressed state,
or with the
probability amplitudes that satisfy $|\alpha
|=0,|\beta |=1$. More
accurately, when the detuning is not swept infinitely
slowly, $\xi >0$,
adiabaticity tends to break down especially in the
neighborhood of
$\delta
=0 $ where the two quasienergies have their closest
approach. The time
evolution of the probability for state $\beta $,
$P(t)\equiv |\beta
(t)|^2$,
often written~\cite{LZSTUFF2} using parabolic cylinder
functions~\cite
{HYPGEO}, can more conveniently be expressed in terms
of the Kummer and
Tricomi confluent hypergeometric functions $_1\!F_1$
and
$U$~\cite{HYPGEO}:
\begin{mathletters}
\begin{eqnarray}
P_{LZ}(\lambda ,t) &=&1-e^{-\pi \lambda }|U(-i\lambda
,\hbox{$1\over2$},-
\hbox{$1\over2$}i\xi t^2)|^2,\hspace{0.35cm} t\le 0,
\label{LinearLZsolA} \\
P_{LZ}(\lambda ,t) &=&1-e^{-\pi \lambda }|
{\frac{2\sqrt{\pi }}{\Gamma
(
\hbox{$1\over2$}-i\lambda )}}\,{}_1\!F_1(-i\lambda
,\hbox{$1\over2$},-
\hbox{$1\over2$}i\xi t^2)  \nonumber \\
&&+\left.U(-i\lambda
,\hbox{$1\over2$},-\hbox{$1\over2$} i\xi
t^2)\right|
^2,\hspace{0.7cm} t\ge 0\,,  \label{LinearLZsolB}
\end{eqnarray}
\label{HYPGEOS}
\end{mathletters}
where $\Gamma $ is the gamma function. At $t=+\infty $
we have the
Landau-Zener result
\begin{equation}
P_{LZ}^\infty (\lambda )\equiv P_{LZ}(\lambda ,\infty
)=1-e^{-4\pi
\lambda
}\,,
\end{equation}
where
\begin{equation}
\lambda ={\frac{\Omega ^2}{2\xi }}  \label{LZPAR}
\end{equation}
is the conventional LZ parameter.

Let us next turn to the nonlinear LZ problem, for which we develop a
series expansion solution. Modern mathematical physics is of course
replete with series expansion techniques. For example, series expansions
have been used to solve an ion in a Paul trap~\cite{PAUL}, the 3-state
Potts~\cite{3PAFT} and Heisenberg~\cite{HEIS} anti-ferromagnetic
models, scattering~\cite{SCATT} and gauge
theories~\cite{GAUGE}, molecular
adsorption~\cite{MOLADS}, black hole
physics~\cite{BHOLE}, as well as
quantum anharmonic~\cite{QAHO,QAHO_DW} and double well
potentials~\cite{QAHO_DW}. Moreover, a series
expansion approach has been
applied to systems with cubic~\cite{CUBE} and
quartic~\cite{QUART}
nonlinearites, which are directly related to the
nonlinear version of the
Landau-Zener problem considered herein.

The equations to solve are a
semiclassical approximation to the photoassociation
problem~\cite{FOOT},
which describes atomic and molecular condensates not
as boson fields
but
as classical fields. We have
\begin{equation}
i\dot{\alpha}={\frac \Omega {\sqrt{2}}}\,\alpha
^{*}\beta ,\qquad i\dot{\beta }={\frac \Omega
{\sqrt{2}}}\,\alpha
^2+\delta \beta \,.  \label{NLEQS}
\end{equation}
Physically, $\alpha$ and
$\beta$ are the probability amplitudes that an atom is part of the atomic
or the molecular condensate. Equations~(\ref {NLEQS}), in fact, preserve
the normalization
$|\alpha |^2+|\beta |^2=1$. The Rabi frequency for
coherent
photoassociation, $\Omega $, may be adjusted by
varying the intensity
of
the driving laser field. The detuning $\delta $, again
swept linearly
as
$\delta =\xi t$, measures the mismatch between the
laser frequency and
the frequency in a transition in which two atoms are
converted into a
molecule. It is controlled by tuning the
photoassociating laser. Transitions to noncondensate
modes~\cite{MULTIMODE,JAV02,85RB} are neglected under the assumption that
the condensate coupling satisfies
$\Omega\ll\hbar\rho^{2/3}/m$~\cite{JAV99,KOS00,JAV02}, where
$\rho$ is the initial atom density and
$m$ is the atomic mass.

For a fixed detuning the system~(\ref{NLEQS}) has
dressed states as
well~
\cite{KOS00}, though they behave quite differently
from the dressed
states
of the linear system~(\ref{LINLZ}). They are of the
form $[\alpha
(t),\beta (t)]=[e^{-i\mu t}\bar{\alpha},e^{-2i\mu
t}\bar{\beta}]$,
where
$\hbar \mu $ is conventionally referred to as chemical
potential.
Depending on the value of the detuning, there may be
as many as three
essentially different dressed states, along with a
multitude of
additional states that can be generated using the
symmetry
transformations of Eqs.~(\ref{NLEQS}). A trivial state
$M$ exists for
all
detunings. It has all atoms as molecules,
$\bar{\beta}=1$, and the
chemical potential equals $\mu =\hbox{$1\over2$}\delta
$. A second state $B$ exists for $\delta /\Omega
<\sqrt{2}$ and has
$\bar{\alpha}=-1$ at  $\delta =-\infty $. For this
state
$\bar{\beta}\rightarrow 1$ and $\mu
\rightarrow \hbox{$1\over2$}\delta $ as $\delta/\Omega
\rightarrow 2$, so that
at $
\delta /\Omega =\sqrt{2}$ the states $M$ and $B$
continuously merge.
The
third state $F$ exists for $\delta /\Omega >-\sqrt{2}$
and turns into
all
atoms, $\bar{\alpha}\rightarrow 1$, in the limit
$\delta \rightarrow
\infty $
.

Suppose now that the system starts as all atoms at a
large negative
detuning, i.e., in state $B$, and that the detuning is
swept slowly
through
the resonance. One expects adiabatic following, so
that the system
should
stay in state $B$ and eventually turn into state $M$,
into molecules.
This
actually is the result both in the corresponding
quantum case addressed
in
Refs.~\cite{JAV99,KOS00}, and from our numerical
trials with Eqs.~(\ref
{NLEQS}). However, the point of resonance where the
degree of the
transfer
of population is mostly determined occurs where the
chemical potentials
are
equal, at $\delta/\Omega=\sqrt2$, which is also
precisely where the
dressed
states merge. The situation is mathematically much
different from the
linear LZ case.

It is therefore gratifying that we have been able to
carry out an
analysis
analogous to the LZ case in our nonlinear system. The
(quite involved)
technical details will be reported elsewhere, here we
only present the
key
idea. Thus, it is possible to combine from
Eqs.~(\ref{NLEQS}) a
third-order
differential equation for the probability for
molecules, $P=|\beta
|^2$,
which reads
\begin{eqnarray}
P^{\prime \prime \prime }+\frac{P^{\prime \prime
}}z+\left[ 1-\frac
1{4z^2}+
\frac{4\lambda }z\left( 1-\frac 32P\right) \right]
P^{\prime } &&
\nonumber
\\
+\frac \lambda {z^2}\left( \frac 12-2P+\frac
32P^2\right) =0\,. &&
\label{3ORD}
\end{eqnarray}
Here the primes refer to derivatives with respect to
the transformed
time
variable $z=\xi t^2/2$. The corresponding equation for
the linear
problem~(
\ref{LINLZ}) is
\begin{eqnarray}
P^{\prime \prime \prime }+\frac{P^{\prime \prime
}}z+\left[ 1-\frac
1{4z^2}+
\frac{4\lambda }z\right] P^{\prime }+\frac \lambda
{z^2}\left[
1-2P\right]
&\equiv &  \nonumber \\
{\cal D}(\lambda ,P) &=&0\,,  \label{3LORD}
\end{eqnarray}
where ${\cal D}$ refers to the rule for forming the
left-hand side of
the
differential equations The solutions to the latter
are, of course,
known; $
P_{LZ}(z)$ of Eqs.~(\ref{HYPGEOS}) is the one with
the appropriate initial conditions at $t=-\infty $.

Now, Eq.~(\ref{3ORD}) may be rearranged to read
\begin{equation}
{\cal D}(\lambda ,2P)=3\lambda \left( \frac{4P^\prime
P}{z}-\frac{P^2}{z^2}
\right) \,.  \label{EQLSMALL}
\end{equation}
In the limit with $\lambda \ll 1$ the LZ transition
probability is
small, and so is presumably the solution to
Eq.~(\ref{3ORD}). In
Eq.~(\ref
{3ORD}) the nonlinear terms, the same ones that make
the right-hand
side of
Eq.~(\ref{EQLSMALL}), are compared with terms that are
much larger. The
nonlinearity makes a ``small'' perturbation. We are
thus lead to
formulate
the Ansatz
\begin{equation}
P(z)=\hbox{$1\over2$}P_{LZ}(\lambda ,z)+\varepsilon
P^{(1)}(z)+\varepsilon
^2P^{(2)}(z)+...\,,  \label{Expansion}
\end{equation}
with $\varepsilon =P_{LZ}^\infty$ may be thought of as
the small
parameter in a perturbative expansion.

The calculations make use of what is essentially the
Green's function
for
the linear differential operator acting on $P$ in
Eq.~(\ref{3LORD}). It
turns out that a formal expansion of the type
(\ref{Expansion}) may be
found, and for a small enough $\lambda $ it even
converges to the
desired
solution of Eq.~(\ref{3ORD}). The leading terms in the
expansion at
infinite
time are
\begin{equation}
P(\infty )=\hbox{$1\over2$}P_{LZ}^{\infty }\left(
1+{\frac{4\lambda
}{\pi }}
P_{LZ}^{\infty }\right) \,.
\end{equation}

Equation~(\ref{3ORD}) may also be rearranged as an
equation for the
atomic
state probability $|\alpha |^2\equiv R=1-P$, and in
this form reads
\begin{equation}
{\cal D}(-\hbox{$1\over2$}\lambda ,R)=3\lambda \left(
-\frac{2R^{\prime
}R}z+
\frac{R^2-1/3}{2z^2}\right) \,.  \label{EQLLARGE}
\end{equation}
In the limit $\lambda \gg 1$ very few atoms will
remain, so one is
tempted
to attempt an Ansatz of the form
\begin{equation}
R(z)=[1-P_{LZ}(\hbox{$1\over2$}\lambda
,z)]+\varepsilon
R^{(1)}(z)+\varepsilon ^2R^{(2)}(z)+...\,.
\label{ExpansionLarge}
\end{equation}
This time the formally small parameter is $\varepsilon
=R(\infty
)=1-P_{LZ}^\infty $. Perhaps surprisingly, this Ansatz
works, too. By
employing essentially the same mathematics as in the
case of the
expansion~(\ref{Expansion}), one finds a formal
expansion that may be
shown to converge for large enough $\lambda $. The
leading terms give
\begin{equation}
P(\infty )=P_{LZ}^\infty \left( {\frac \lambda
2}\right) \left[
1-{\frac 1{
3\pi \lambda }}P_{LZ}^\infty \left( {\frac \lambda
2}\right) \right]
\,.
\label{PlargeLAMBDA}
\end{equation}

The most notable qualitative difference from the
linear case occurs in
the
limit of slow sweep of the frequency, $\lambda\gg1$.
The probability
for no
transition, $|\alpha(\infty)|^2 = 1-P(\infty)$,
behaves in the linear
case
like $\exp(-4\pi\lambda)$, but in the nonlinear case
as
$(3\pi\lambda)^{-1}$.

\begin{figure}
\centering
\epsfig{file=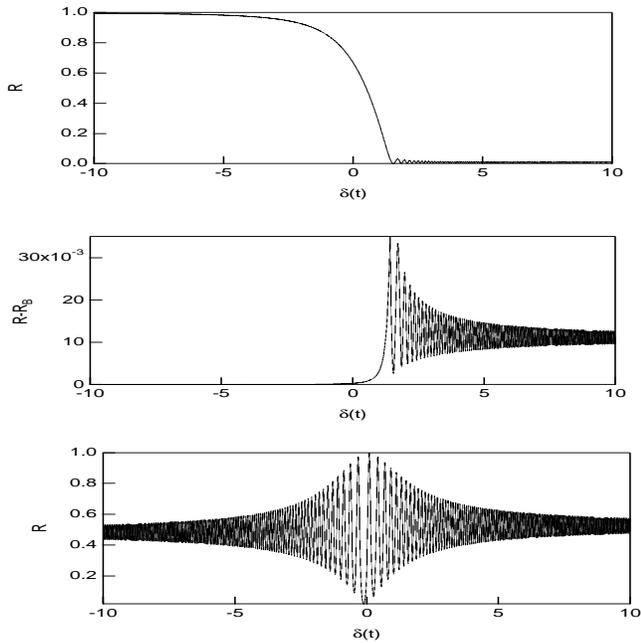,width=8.5cm,height=8.5cm}
\caption{ (a) Probability of remaining in the atomic condensate
$R=1-P$ as a function of running detuning $\delta=\xi t$, with $\xi =
0.01\,\Omega^2$. The system is started at a large negative
time in the dressed eigenstate $B$, and arrives adiabatically from the
left to the range of $\delta$ shown in the figure.
(b) Difference between the atomic probability $R$ and the
adiabatic probability $R_B$ of the dressed state $B$ for the same data
as in panel (a). We set $R_B=0$ for $\delta>\protect\sqrt{2}\,\Omega$,
when the state $B$ has merged with the all-molecules state $M$.
(c) Same as panel~(a), except that the system was started in a
superposition with $\alpha=\beta=1/\protect\sqrt{2}$.}
\label{ADF}
\end{figure}

We have solved Eqs.~(\ref{NLEQS}) also numerically
using several
methods. An example is shown in Fig.~\ref{ADF}(a),
where we plot the
atomic
probability $R=1-P = |\alpha|^2$ as a function of the
running detuning
$
\delta=\xi t$. In this example we initially fix a
negative detuning
that lies outside the range of the graph, and start
the system in the
corresponding dressed state $B$. We then slowly
increase the rate of
the
sweep of the detuning until it reaches
$\xi=0.1\,\Omega^2$, still far
to
the left of the detuning axis in Fig.~\ref{ADF}(a).
Upon arrival to the
range plotted in Fig.~\ref{ADF}(a), the detuning is
swept at this
constant
rate
$\xi$, and the occupation probability is within
$10^{-8}$ of the
occupation probability for the dressed state $B$. At
this
point adiabaticity still prevails. Especially around
$\delta/\Omega\simeq\sqrt2$, though, adiabaticity
breaks down. To
demonstrate, we show in Fig.\ref{ADF}(b) the
difference between the
actual occupation probability of the atomic state and
the prediction
from
the dressed state $B$. Typical of both linear and nonlinear numerical
solutions, oscillations arise because the system
does not perfectly follow the ground state, and filtering out these
oscillations accurately is a major tour de force.
Nonetheless, it is easy to see the $\lambda^{-1}$ dependence of the
residual atomic
probability on the sweep rate numerically.

While these aspects of the computations are somewhat
trivial,
computations also provide new insights. The linear LZ
problem is
completely symmetric in the two states. Whether the
system starts in
state $\alpha$ or $\beta$, the probability for the
transition to the
other state is the same. This cannot hold in the
nonlinear case; if
$\alpha=0$ at some time, then  by Eqs.~(\ref{NLEQS})
it will remain so
forevermore. Also, given that the states $B$ and $M$
merge at
$\delta=\sqrt{2}\,\Omega$, one might expect that there
is some
preponderance to transitions to a molecular state.
This does not
seem to be the case. In Fig.~\ref{ADF}(c) we resort to
the same scheme
as
in Fig.~\ref{ADF}(a), except that we start the system
in the
superposition with
$\alpha=\beta={\frac{1}{\sqrt{2}}}$. The
probabilities
oscillate violently around $\delta=0$, but eventually
settle in the
neighborhood of half and half again.

We now briefly consider explicit numbers for $^{23}$Na and
$^{87}$Rb condensates. First is the validity
of the two-mode model~(\ref{NLEQS}), which again hinges on the condition
$\Omega\ll\hbar\rho^{2/3}/m$. The condensate coupling is~\cite{KOS00}:
$\Omega=[(\lambdabar^3\rho)(I/I_0)]^{1/2}$, where $2\pi\lambdabar$ is the
wavelength of the photoassociating light and $I_0$ is the chracteristic
intensity of a photoassociation transition with binding energy
$\sim 1\,\text{cm}^{-1}$. Given the $^{23}$Na ($^{87}$Rb) characteristic
intensity $I_0=0.47\,\text{W/cm}^2$
($I_0=0.07\,\text{W/cm}^2$)~\cite{KOS00}, and a typical
 BEC density $\rho=5\times 10^{14}\,\text{cm}^{-3}$,
the two-mode model is valid for $I\ll 50\,\text{mW/cm}^2$
($I\ll 10\,\text{mW/cm}^2$). The linear rate dependence for the
transition probability should manifest for slow sweep rates, i.e.,
$\xi\sim0.01\,\Omega^2$. For intensities satisfying the above
relation, and typical BEC densities, this
translates into a $^{23}$Na ($^{87}$Rb) detuning-sweep rate
$\xi\sim 200\times 2\pi\,\text{Hz}$
($\xi\sim 50\times 2\pi\,\text{Hz}$).

In sum, we have studied a nonlinear variant of the
Landau-Zener
problem.
Though our specific example was about photoassociation
of a
Bose-Einstein
condensate, the problem is generic in classical and
bosonic field
theories
with a cubic nonlinearity. While the basic
adiabaticity argument still
works, the structure of the adiabatic or ``dressed''
states in the
nonlinear
system is much different from its counterpart in
standard linear
quantum
mechanics. When the levels cross slowly, the
probability for the
failure of
adiabaticity turns out to scale linearly with the rate
at which the
levels cross, as opposed to the exponential behavior
of the linear
problem. This prediction could be tested with existing $^{23}$Na
and
$^{87}$Rb experimental setups.

\begin{references}
\bibitem{FOOT}  We take the semiclassical equations
from
Ref.~\cite{KOS00}
and scale the amplitude of the molecular field by
$\sqrt{2}$.
\end{references}
\end{document}